\documentclass[utf8]{frontiersSCNS} 

\usepackage{url,hyperref,lineno,microtype}
\usepackage[onehalfspacing]{setspace}

\newcommand{\utwi}[1]{\mbox{\boldmath $ #1$}}

\newcommand{\bp}{{\utwi{p}}}

\newcommand{\bs}{{\utwi{s}}}

\newcommand{\bx}{{\utwi{x}}}

\def\keyFont{\fontsize{8}{11}\helveticabold }

\def\firstAuthorLast{Terebus {et~al.}} 
\def\Authors{Anna Terebus$^{1,2,\dagger}$, Farid Manuchehrfar$^{1,\dagger}$, Youfang Cao$^{1,3}$, and Jie Liang$^{1,*}$}
\def\Address{$^{\dagger}$Authors have equal contribution $^{1}$Center for Bioinformatics and Quantitative Biology, Richard and Loan Hill Department of Bioengineering, University of Illinois at Chicago, Chicago IL, 60607, USA;
$^{2}$Constellation, Baltimore, MD, USA;
$^3$Merck \& Co., Inc., Kenilworth, NJ, USA.
}

\def\corrAuthor{Jie Liang}

\def\corrEmail{jliang@uic.edu}

\begin{document}

\onecolumn


\title[Stochastic multimodality of Feed-Forward Loops]
{Exact Probability Landscapes of Stochastic Phenotype Switching in Feed-Forward Loops: Phase Diagrams of Multimodality} 
\author[\firstAuthorLast ]{\Authors} 
\address{} 
\correspondance{} 
\extraAuth{}
\maketitle

\begin{abstract}
    
Feed-forward loops (FFLs) are among the most ubiquitously found  motifs of reaction networks in nature.  However, little is known about their stochastic behavior and the variety of network phenotypes they can exhibit. In this study, we provide full characterizations of the properties of stochastic  multimodality of FFLs, and how switching between different network phenotypes are controlled. 
We have computed the exact steady state probability landscapes of all eight types of coherent and incoherent FFLs using the finite-butter ACME algorithm, and quantified the exact topological features of their high-dimensional probability landscapes using persistent homology. 
Through analysis of the degree of multimodality for each of a set of 10,812 probability landscapes, where each landscape resides over $10^5$--$10^6$ microstates, we have constructed comprehensive phase diagrams of all relevant behavior of FFL multimodality over broad ranges of input and regulation intensities, as well as different regimes of promoter binding dynamics.  In addition, we have quantified the topological sensitivity of the multimodality of the landscapes to regulation intensities.
Our results show that with slow binding and unbinding dynamics of transcription factor to promoter,  FFLs exhibit strong stochastic behavior that is very different from what would be inferred from deterministic models. 
In addition, input intensity play major roles in the phenotypes of FFLs: At weak input intensity, FFL exhibit monomodality, but strong input intensity may result in up to 6 stable phenotypes. 
Furthermore, we found that gene duplication can  enlarge stable regions of specific multimodalities and enrich the phenotypic diversity of FFL networks, providing means for cells towards better adaptation to changing environment.
Our results are directly applicable to analysis of behavior of 
FFLs in biological processes such as stem cell differentiation and for design of synthetic networks when certain phenotypic behavior is desired.

\tiny
 \keyFont{ \section{Keywords:} System Biology, Feed Forward Loop, Gene Regulatory Network, Network Motif, Stochastic Reaction Network, Persistent Homology, Finite Buffer Algorithm, ACME Algorithm, Topological Data Analysis} 

\end{abstract}

\section{Introduction}

Cells with the same genetic make-ups can exhibit a variety of different behavior. They can also switch between these different phenotypes stochastically. This phenomenon has been observed in bacteria, yeast, and mammals such as neural cells~\citep{acar2005enhancement,choi2008stochastic,Guo2009,
gupta2011stochastic}.  The ability to exhibit multiple phenotypes and switching between them
 is the foundation of cellular fate decision~\citep{schultz2007molecular,cao2010probability,  chunhe2019}, 
stem cell differentiation~\citep{feng2012new, papatsenko2015single, chunhe2019-2}, and tumor formation~\citep{huang2009cancer, shiraishi2010large}.

Cells exhibiting different phenotypes have different patterns of gene expression.
Single-cell studies demonstrated that isogenic cells can exhibit different modes of gene
expression~\citep{shalek2013single}, indicating that  
distinct phenotypes are encoded in the wiring of the genetic regulatory networks. 
This phenomenon of epigenetic control of bimodality in gene expression by network architecture is well known and has been extensively studied in earlier works of phage-lambda~\citep{Arkin1633,Ptashne2004,ZhuFIG2004,ZhuJBCB2004,Cao18445}. 
Understanding multimodality in gene regulatory networks and its control mechanism can provide valuable insight into how different cellular phenotypes arises and how cellular programming and reprogramming proceed~\citep{lu2007phenotypic}. 
Much of current knowledge of multimodality is  derived from analysis of networks with
feedback loops or cooperative interactions~\citep{siegal2009capacity}.  
However, recent studies suggest that multimodality and phenotype switching can also arise from slow promoter binding, which may result in
distinct protein expression levels of long durations~\citep{feng2012new,thomas2014phenotypic, chen2015distinguishing, duncan2015noise,terebus2019sensitivities}. Nevertheless,
the nature and extent of this type of bimodality is not well understood. 

In this work we study the  network modules of  feed-forward loops (FFLs) and characterize the stochastic nature of their multimodalities.
FFLs are one of the most prevalent three-node network motifs in  nature~\citep{alon2006introduction} and play important regulatory roles~\citep{lee2002transcriptional,shen2002network, boyer2005core,mangan2006incoherent,tsang2007microrna, ma2009defining, sorrells2015making}. 
They appear in stem cell pluripotency networks~\citep{boyer2005core,sorrells2015making, papatsenko2015single},
microRNA regulation networks~\citep{tsang2007microrna, re2009genome, ivey2010micrornas}, and cancer networks~\citep{re2009genome}.
The behavior of FFLs has been studied extensively using deterministic ODE models.
These studies revealed important functions of feed-forward loops in signal processing, including sign-sensitive acceleration and delay pulse generation functions, and increased cooperativity~\citep{mangan2003structure, ma2009defining}. 
FFLs are also found to be capable of maintaining robust adaptation~\citep{franccois2008case, ma2009defining} 
and detecting ``fold-changes''~\citep{goentoro2009incoherent}.

However, analysis based on ODEs
is limited in its ability to characterize probabilistic events, as they do not  capture bimodality in gene expression that arises from slow promoter.
binding~\citep{vellela2009}.
The stochastic behavior of FFLs is not well characterized: 
Basic properties such as the number of different phenotypes FFLs are capable of exhibiting,  the conditions required for their emergency, their relative prominence, and
the sensitivity of different phenotypes to perturbations are not known.

Our stochastic FFL models are based on processes of
 Stochastic Chemical Kinetics (SCK), which provides a general framework for understanding the stochastic behavior of reaction networks.
Quantitative SCK modeling can uncover  different  network phenotypes, the conditions for their occurrence, and  the nature of the prominence of the stability peaks.
However, analysis of stochastic networks is challenging. First, models based on stochastic differential equations  such as Fokker-Planck and Lagenvin models may be inadequate due to their  Gaussian approximations.
This is further compounded by the limited number of simulation trajectories that can be generated.  These difficulties are reflected in the reported
failure of a Fokker-Planck model in accounting for multimodality 
in the simple network model of single self-regulating gene at certain reaction rates~\citep{duncan2015noise}.
Second, the widely used Stochastic Simulation Algorithm (Gillespie simulations) can generate SCK trajectories~\citep{gillespie1977exact}, but are challenged in capturing rare events and in computing efficiency. There are also difficulties in assessing convergency and in estimating computational errors~\citep{cao2013}. 
Third, even if the probabilistic landscape can be accurately reconstructed with acceptable accuracy, detecting topological features such as peaks in high-dimensional probability landscapes and assessing their objectively prominence at large scale remains an unsolved problem.

In this study, we characterize the stochastic behavior of FFLs using models based on SCK processes. Our approach is solving the underlying
discrete Chemical Master Equation (dCME) using the ACME (Accurate Chemical Master Equation) algorithm~\citep{cao2016state, caoaccurate}, we obtain the exact probability landscapes of all 8 varieties of FFLs.

Aided by the computational efficiency of ACME,
we are able to explore the behavior FFLs under broad conditions of synthesis, degradation, 
 binding and unbinding rates of transcription factors genes binding. 
Furthermore, we analyze the topological features  of the exactly constructed high-dimensional probability landscapes using persistent homology, so the number of probability peaks and the prominence measured by their persistence are quantified objectively.  
These techniques allow us to examine details of  the number of possible phenotypic states at different conditions, as well as the ranges of conditions enabling phenotypic  switching. 
With broad exploration of model parameter space, we are able to construct detailed phase diagram of multimodalities under different conditions. 

Our results reveal how FFL network behaves differently under varying strengths of regulations intensities and the input.  
In addition,  we characterize quantitatively the effects of duplication of genes in the FFL network modules. We show gene duplication can affect significantly the diversity of multimodality, and can enlarge monomodal regions so FFLs can have robust phenotypes. 
The results we obtained can be useful for analysis of phenotypic switching in biological networks containing the feed-forward loop modules.  They can also be used for construction of synthetic networks with the goal of generating certain desired phenotypic behavior.

\section {Models and Methods}
\subsection{Architecture and types of feed-forward loop network modules}

\begin{figure}[h!]
\centering
\includegraphics[width=1\linewidth]{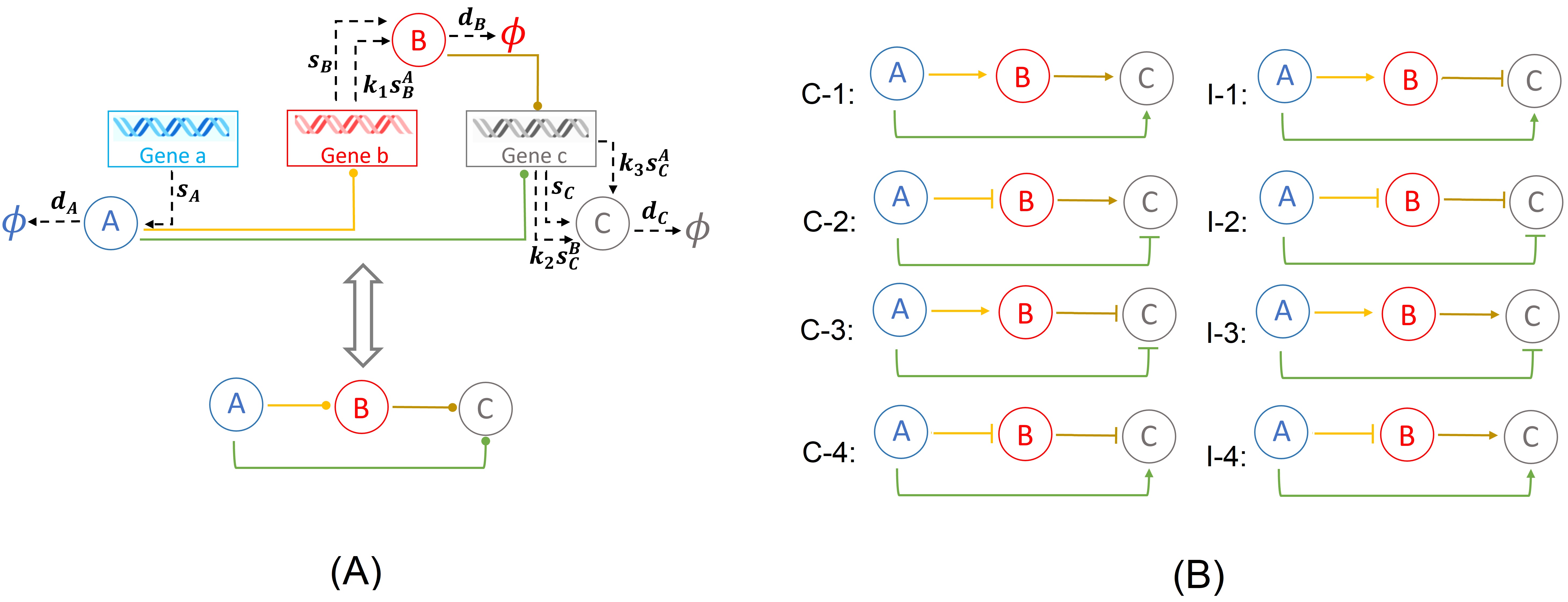}
\caption{Representation and the types of FFL network: 
(A) General wiring and corresponding 3-node schematic representation of an FFL module containing three genes $a$, $b$, $c$ 
expressing three proteins $A$, $B$, $C$. Protein $A$ regulates the expressions of genes $b$ and $c$ 
through binding to their promoters. Protein $B$ regulates the expression of gene $c$ through promoter binding. (B) The FFL modules can be classified into  eight different types.  Coherent/incoherent FFLs are on the left/right, respectively. }
\label{fig:Fig_loop}
\end{figure}

{\bf Overview.}
FFLs consists of three nodes representing three genes, each expresses a different protein product~(Fig.~\ref{fig:Fig_loop}A).  
An FFL regulates the network output from the left input  node towards the right output node via
two paths; the direct path from the left node to the right  node, and
the indirect path from the left to 
the right  node via an intermediate buffer node.
As each of the three regulations can be either up- or down-regulation,  there are altogether $2^3=8$ types of feed-forward loop.

{\bf Network architecture.} 
Specifically, we denote the three genes of an FFL module as $a$, $b$, and $c$, which expresses protein products $A$, $B$, and $C$ at constant synthesis rate  of $s_A$, $s_B$ and $s_C$, respectively (Fig.~\ref{fig:Fig_loop}A). Proteins $A$, $B$, and $C$ are degraded
at rate $d_A$, $d_B$ and $d_C$, respectively.  
Both proteins
$A$ and $B$ function as transcription factors and can
bind competitively to  
the promoter of gene $c$ and regulates its expression. As the  promoter of gene $c$ can bind to either protein $A$ or $B$, but not both, this type of regulation is known as the ``OR'' gate.
In addition, protein $A$ can bind to the promoter of gene $b$ and regulate its expression.
Specifically,
protein $A$ can bind to the promoter of gene $c$ at rate $r_c^A$ to form complex $cA$, which dissociates at rate
 $f_c^A$.   $cA$ expresses protein $C$ at a rate  $k_3$-fold over the basal rate of $s_C$.
Similarly, protein $B$ can  bind to the promoter of gene $c$ at rate $r_c^B$ to form complex $cB$, which dissociates at rate $f_c^B$.  
 $cB$ expresses protein $C$ at a rate  $k_2$-fold over the basal rate of $s_C$.
Furthermore, protein $A$ binds to the promoter of gene $b$ at rate $r_b^A$ to form gene-protein complex $bA$, which
dissociate at rate $f_b^A$. 
Upon binding protein $A$,  $bA$ expresses protein $B$ at a rate  $k_1$-fold over the basal rate of $s_B$.

The biochemical reactions of our FFL model are summarized below: 
$$ b + A \stackrel{r_b^{A}}{\rightarrow} bA; \quad bA \stackrel{f_b^{A}}{\rightarrow} b + A ;\\$$
$$ c + A \stackrel{r_c^{A}}{\rightarrow} cA; \quad cA \stackrel{f_c^{A}}{\rightarrow} c + A;\\$$
$$ c + B \stackrel{r_c^{B}}{\rightarrow} cB; \quad cB \stackrel{f_c^{B}}{\rightarrow} c + B;\\$$
$$ a \stackrel{s_{A}} {\rightarrow} a + A; \quad A \stackrel{d_{A}} {\rightarrow} \emptyset; \\$$ 
$$ b \stackrel{s_{B}} {\rightarrow} b + B; \quad bA \stackrel{s_{B}*k_1} {\rightarrow} bA + B; \quad B \stackrel{d_{B}=1} {\rightarrow} \emptyset; \\$$
$$ c \stackrel{s_{C}} {\rightarrow} c + C; \quad cB \stackrel{s_{C}*k_2} {\rightarrow} cB + C; \quad cA \stackrel{s_{C}*k_3} {\rightarrow} cA + C; \quad C \stackrel{d_{C}} {\rightarrow} \emptyset. \\$$
Here we set  $r_b^{A}=r_c^{A}=r_c^{B}=0.005~s^{-1}$, $f_b^{A}=f_c^{A}=f_c^{B}=0.1~s^{-1}$, $d_A=d_B=d_C=1~s^{-1}$, and $s_{A}=s_{B}=s_{C}=10~s^{-1}$.
All reaction rate constants are of the unit $s^{-1}$, while coefficients $k_1, k_2,$ and $k_3$ are ratio of reaction rates and therefore unitless. The ratios $k_1, k_2$ and $k_3$ can take different values so the network represents different types of feed-forward loops.

{\bf Types of feed-forward loop modules.}
Depending on the nature of the regulations, namely, whether each of regulation intensities  $k_1$, $k_2$, and $k_3$ is   $\ge 1$ (activating) or $<1$ (inhibiting),
there are $2^3 = 8$ types of feed-forward loops.
These FFLs 
are classified into two classes, the \emph{coherent feed-forward loops}
and the \emph{incoherent feed-forward loops} (Fig.~\ref{fig:Fig_loop}B)~\citep{alon2006introduction}.
A feed-forward loop is termed \emph{coherent}  ($C_1$, $C_2$, $C_3$, $C_4$ on the Fig.~\ref{fig:Fig_loop} (B)),
if the direct effect of protein $A$ on the gene $c$ has the same sign (positive or negative) as its net
indirect effect through protein $B$. 
Taking the FFL model $C_1$ (Fig.~\ref{fig:Fig_loop}B) as an example,
protein $A$ activates gene $b$, and protein $B$ activates gene $c$, with an overall effect of ``activation''.
 At  the same time,  the direct effect of product of gene $a$ protein $A$ is also
 activation of gene $c$.  Therefore, $C_1$ is a coherent FFL.
When the sign of the
indirect path of the regulation is opposite to that of the direct path, we have  \emph{incoherent} FFLs  ($I_1$, $I_2$, $I_3$, $I_4$ in Fig.~\ref{fig:Fig_loop}B).
Takeing the FFL model $I_1$ as an example, the effect of the 
 direct path is positive, but the overall effect of the indirect path is negative.
As can be seen from Fig.~\ref{fig:Fig_loop}B, all incoherent
FFLs have an odd number of edges of inhibition.

{\bf Model parameters.}
In order to explore broadly the behavior of all types of FFLs,
we construct FFL models over the parameter space of
a wide range of possible combinations
of  $k_1$, $k_2$, and $k_3$, representing  all  8  types  of  FFLs.
The  regulation intensity are set to values based on values reported in~\citep{BU2016189, Tej_2019}. We then altered the regulation intensities by about 10 folds to study the general behavior of different types of FFLs at the steady state.
We take parameter values of $k_1 \in \{0.025, 0.1, 0.4, 0.8, 1.5, 2.1, 2.4, 3.0 \}$, $k_2 \in [0.025,\, 5.0]$ with step size of $0.25$, $k_3 \in [0.025,\, 5.0]$ with step size of $0.25$. In addition, for the input intensity, the values are selected based on the analysis of abundance pattern reported in~\citep{momin2020}. We take  $s_A \in { \{3.0, 10.0 \}} s^{-1}$, $r_c^A$ and $r_c^B \in {\{ 0.5, 2, 8, 16 \}}s^{-1}$ for one and two copies of genes $b$ and $c$.
Details of the relationship of FFL types with $k_1$, $k_2$, and $k_3$ are listed in Table~\ref{tab:1}.
Over this parameter space, 
 we study the behavior of all 8 types of FFLs.
Overall,
we constructed a total of 10,812 examples of FFLs and computed the steady state probability landscape for each of them. 

\,

\subsection{Computing probability landscape using ACME}

{\bf Exact computation of probability landscape of FFLs.}
Consider a well mixed system of reaction with constant volume and temperature. This system has $n$ species $X_i$, $i=1, 2, \cdots, n$, in which each particle can participate in $m$ reactions $R_k$, $k=1, 2, \cdots, m$. A microstate of the system at time $t$, $\bx(t)$ is a column vector representing the copy number of species: $\bx(t)=(x_1(t), x_2(t), \cdots, x_n(t))^T$, 
where the values of copy numbers are non-negative integers. The state space $\Omega$ of the system includes all the possible microstate of the system from $t=0$ to infinity, $\Omega=\{ \bx(t)|  t \in [0,\infty ) \}$. In this study, the size of the state space is $|\Omega|=657,900$ when genes $b$ and $c$ are single-copy, and $|\Omega|=686,052$ and $1,289,656$ when there are two copies of gene $b$ and $c$, respectively.

The reaction $R_k$ of the system takes the form of 
$$R_k: c_{1_k}X_1 + c_{2_k}X_2 + \cdots +c_{n_k}x_n \xrightarrow{r_k} c'_{1_k}X_1 + c'_{2_k}X_2 + \cdots +c'_{n_k}x_n  $$ 
which brings the system from a microstate $\bx$ to a new microstate $\bx+\bs_k$, where $\bs_k$ is the stoichiometry vector and is defined as 
$$\bs_k= (c'_{1_k}-c_{1_k}, c'_{2_k}-c_{2_k}, \cdots, c'_{2_k}-c_{2_k}).$$

In a well mixed system, the propensity function of reaction $k$, $A_k(\bx)$ is given by the product of the intrinsic reaction rate constant $r_k$ and possible combinations of the relevant reactants in the current state $\bx$. 
$$ A_k(\bx) = r_k \prod_{l=1}^{n} {x_l \choose c_{l_k}}$$
With the above definitions, the discrete Chemical Master Equation (dCME) of a network model of the SCK processes consists of a set of linear ordinary differential equations defining the changes in the probability landscape over time at each microstate $\bx$. Denote the probability of the system at a specific microstate $\bx$ at time $t$ as $p(\bx, t) \in \mathbb{R}_{[0,1]}$, the probability landscape of the system over the whole state space $\Omega$ as $\bp(t)=\{p(\bx(t))| \bx(t)\in\Omega\}$, the dCME of the system can be written as the general form of 
$$
\frac{dp(\bx,t)}{dt}=\sum_{k=1}^{m} [A_k(\bx-\bx_k)p(\bx-s_k,t)-A_k(\bx)p(\bx,t)],
$$
where $\bx$ and $\bx-s_k$ $\in \Omega$.

The steady state probability landscapes is obtained by solving the dCME directly.
The exact solution is made possible by using the 
the  ACME algorithm~\citep{caoaccurate, cao2016state}. The ACME algorithm eliminates potential problems due to inadequate sampling, where rare events of very low probability is difficult to estimate using techniques such as the stochastic simulation algorithm (SSA)~\citep{gillespie1977exact,kuwahara2008, Daigle2011, cao2013}. 

\

\subsection{Identification of multimodality by persistent homology}

Despite its simple architecture, FFLs have a 9-dimensional probability landscape: There are three genes ($a$, $b$, and $c$), three proteins ($A$, $B$, and $C$), and three bound genes $bA$, $cA$, and $cB$ (\textit{i.e.},  gene $b$ bound to protein $A$, gene $c$ bound to either protein $A$ or
protein $B$). 
Because of the high dimensionality,
it is challenging to characterize the topological structures of their probability landscapes; Restricting networks to only ``on'' and ``of'' state separately makes it difficult to gain insight into the overall behavior of the network. 

However, quantifying mutlistability at the steady-state is challenging.
Finding peak states by examining distinct local maxima  is equivalent to locating hypercubes that are critical points of Morse index of $d$ in the $d$-dimension state space. 
While local maxima may be identified by comparing its probability value with those of all of its neighbors, all peaks regardless their prominence will be identified.  As numerical calculation may introduce small errors, peaks of tiny magnitude will be included. It is non-trivial to decide on a proper threshold to filter them out.
Persistent homology provides an exact method for identifying the prominent probability peaks.

There have been studies that analyzing $d$-dimensional probability landscape by examining its projection onto 1-d or 2-d subspaces (\textit{e.g.}, $2$-d heatmaps or contour plots)~\citep{BU2016189, Anupam2021}.
However,   projected probability surface on lower dimensional space often no longer reflect the topology of the original space, with results and interpretations likely erroneous or misleading~\citep{OUR_TPS_2021}.

Persistent homology provides a powerful  method  that can characterize topological features of high dimensional probability landscapes~\citep{Edelsbrunner2002,carlsson2009topology}.  Here we use newly developed cubic complex algorithm to compute homology groups~\citep{Tian-topo-preprint} and quantitatively assess the exact topology of the 9-dimensional probability landscape.

{\bf Homology groups.}
We use homology groups from algebraic 
topology to characterize the probability landscape.
Homology group provides an unambiguous and quantitative description on how a space is 
connected. It returns a set of algebraic
groups describing topological features of holes of various dimension in the space.
 The rank of each $i$-th groups  counts the number of linearly independent holes in the 
 corresponding $i$-th dimension. For example, $\text{Rank}(H_0)$ counts the number of 
 connected components ($0$-th dimensional holes).

\textbf{Persistent homolgy}. 
Persistent homology  
measures the importance of these 
topological features~\citep{Edelsbrunner2002}, and has been applied in studies of chemical compounds and biomolecules~\citep{xia2014persistent,xia2015multidimensional, xia2015persistent}.
Here we focus on the topological features of probability peaks, including their appearance 
and disappearance. They are measured by persistent homology of the $0$-th homology 
group.
Specifically, we take the probability $p(\bx)$ as a height function, and construct a sequence of topological spaces using thresholds $\{r_i\}$ for $p(\bx)$:

\begin{equation}
	1 = r_0 > r_1 > r_2 > \cdots > r_{i_{n-1}} > r_{i_n} = 0,
	\label{eqn:seq}
\end{equation}

The superlevel sets $\{{X_i}\}$ has  ${X_i}=\{\bx\in {X}|p(\bx)\ge r_i\}$, which corresponds to the threshold  $r_i$. The sequence  $\{{X_i}\}$ gives a sequence of subspaces, which is called {\it filtration}:

\begin{equation}
	\varnothing \equiv X_{i_0} \subset {X_{i_1}} \subset {X_{i_2}} \subset \cdots
	\subset {X_{i_{n-1}}} \subset {X_{i_n}} 
	\equiv \Omega,
	\label{eqn:filtration}
\end{equation}

As the threshold changes, the peak of a probability landscape emerges from the sea-level at a specific threshold, which is the {\it birth time} of the corresponding 0-homology group in the filtration.  It disappears
as an independent component when merged with a prior peak at a particular threshold, which is called the {\it death time}.  When the sea-level recedes to the ground level at $p(\bx)=0$, only the first peak remains.

{\bf Persistent diagram  of multimodality in probability landscape.}
We keep track of the probability peaks by recording the birth and death times of their corresponding $0$-homology groups throughout the filtration. 
This relationship is  depicted  by the two-dimensional {\it persistent diagram}.

For the $i$-th probability peak, when the threshold  $r$ reaches the value $r_b(i)$, the probability peak appears.  We call this value the {\it birth probability}\/ $p_b(i)=r_b(i)$ of peak $i$. When the threshold $r$ is lowered to a value $r_d(i)$, this peak is merged to an existing peak.  We call this value the {\it death probability}\/ $p_d(i)=r_d(i)$ of peak $i$.  The {\it persistence}\/ of peak $i$ is defined as:
\begin{equation}
    \mathrm{pers}(i) \equiv p_b(i) - p_d(i).
    \label{eqn:persistence}
\end{equation}
The {\it persistent diagram}\/ plots
 peak $i$ using the  birth probability $p_b(i)$ as the $x$-coordinate and the death probability $p_d(i)$ as the $y$ coordinate.
The number of dots on the persistent diagram corresponds to the number of probability peaks. Those that are further off the diagonals are the more prominent probability peaks as their persistences are larger.

\section {Results}

\subsection{Multimodality and persistent homology of FFLs.}

\begin{figure}[h!]
\centering
\includegraphics[width=0.7\linewidth]{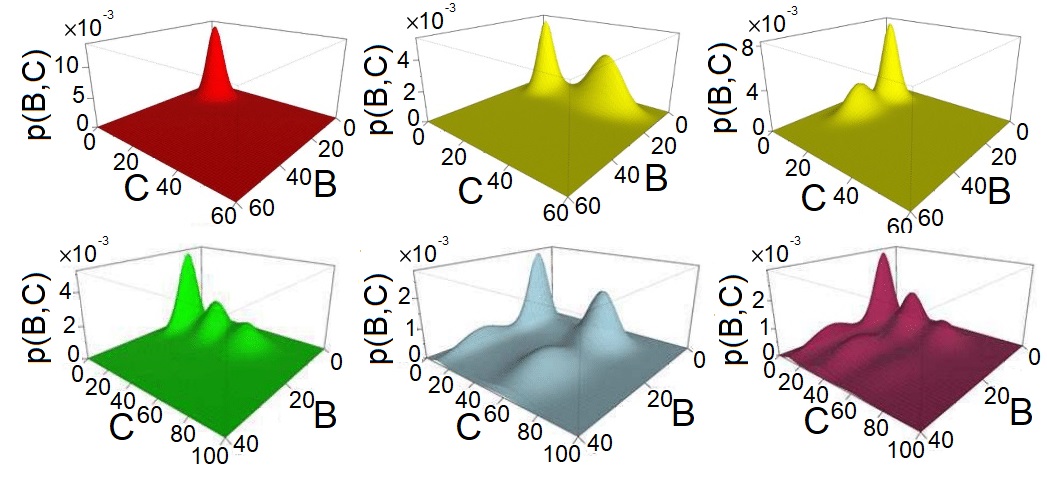}
\caption
{ Examples of multimodality exhibited by Feed Forward Loop (FFL) network motifs. The steady state probability landscape can exhibit up to 6 different multimodes. The illustrative examples are:
1 peak (red),  coherent FFL of type C1 when $k_1  = 1.2$, $k_2 = 1.2$, and $k_3 = 1.2$; 2 peaks (yellow), either in protein $B$ with coherent FFL of type C1, where $k_1  = 3.0$, $k_2 = 1.2$, and $k_3 = 1.2$,
or in protein $C$ with coherent FFL of type C1,   where $k_1  = 1.2$, $k_2 = 6.0$, and $k_3 = 6.0$;
3 peaks (green),  coherent FFL of type C1, where $k_1  = 1.2$, $k_2 = 6.0$, and $k_3 = 3.6$; 
4 peaks (light-blue), coherent FFL of type C1 exhibits two peaks for protein $B$ and two peaks for protein $C$, where $k_1  = 3.0$, $k_2 = 6.0$, and $k_3 = 6.0$;
and
6 peaks (purple),  coherent FFL of type C1 exhibit two peaks for $B$ and three peaks for $C$, where $k_1  = 3.0$, $k_2 = 6.0$, and $k_3 = 3.6$.
}
\label{fig:Fig_examples}
\end{figure}

For each FFL network, we first compute its probability landscapes 
$p=p(x_A,$\,$ x_B,$\,$ x_C,$\,$ x_a,$\,$ x_b,$\,$ x_c, $\,$ x_{bA},$\,$ x_{cA},$\, $ x_{cB})$  at the steady state under various conditions of model parameters. Here $x_A$, $x_B$, and $x_C$ are copy numbers of proteins $A$, $B$, and $C$, respectively; $x_a$, $x_b$, and $x_c$ are copy numbers of genes $a$, $b$, and $c$, respectively; $x_{bA}$ and $x_{cA}$ are copy numbers of genes $b$ and $c$ bound by protein $A$; $x_{cB}$ is the copy number of gene $c$ bound by protein $B$.

Our results show that the 8 types of FFLs can exhibit 
up to six different phenotypes of mono- and  multimodality  at different conditions in  the parameter spaces we investigated.  Illustration of these six different types of multimodality are shown in Fig.~\ref{fig:Fig_examples}. 

\begin{figure}[h!]
\centering
\includegraphics[width=0.7\linewidth]{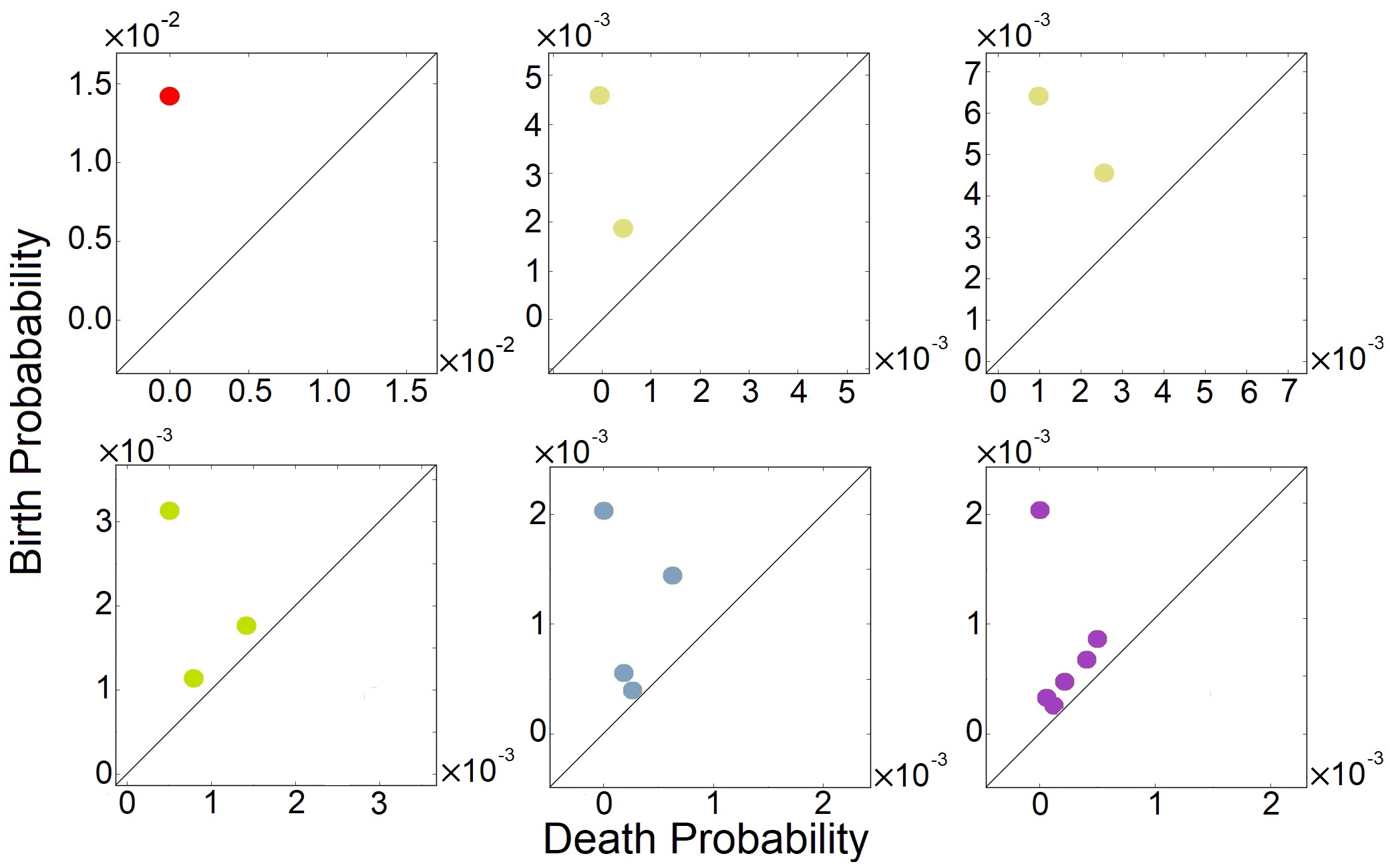}
\caption
{Persistent diagrams (PDs)  of FFL network modules of Fig~\ref{fig:Fig_examples} exhibiting different multimodalities.
(Red) the probability landscape  with monomodality. 
(Yellow) these two PDs  depict the two steady state landscapes exhibiting bimodality.
(Green, light blue, and purple) these three PDs depict the landscape exhibiting tri-modality,
4-modality, and 6-modality, respectively.
}
\label{fig:PD}
\end{figure}

We further computed their 0-th homology groups  at varying sea level of probability.  
The number of peaks, the birth and death probability associated with each peak for examples in Fig~\ref{fig:Fig_examples} are
shown in the persistent diagrams of 
Fig.~\ref{fig:PD}.

\ 

\subsubsection{Behavior of FFLs from stochastic models differ from deterministic ODE models}

The behavior of FFL network modules revealed from our stochastic models are fundamentally different from that of deterministic models of ordinary differential equations (ODEs). ODE models are based on kinetics of law of mass-action  and are used to
calculate the mean concentrations of $A$, $B$, and $C$ at equilibrium state.
However, they do not provide accurate pictures on the degree of multimodality.
For example, the steady state ODE solutions  with respect to different gene occupancy  for mass action kinetics show that there are at most six phenotypic states (see supplementary information for more details).
However, as there are no probabilistic considerations, 
conclusions drawn from ODE models can be problematic.

\begin{figure}[h!]
\centering
\includegraphics[scale=0.54]{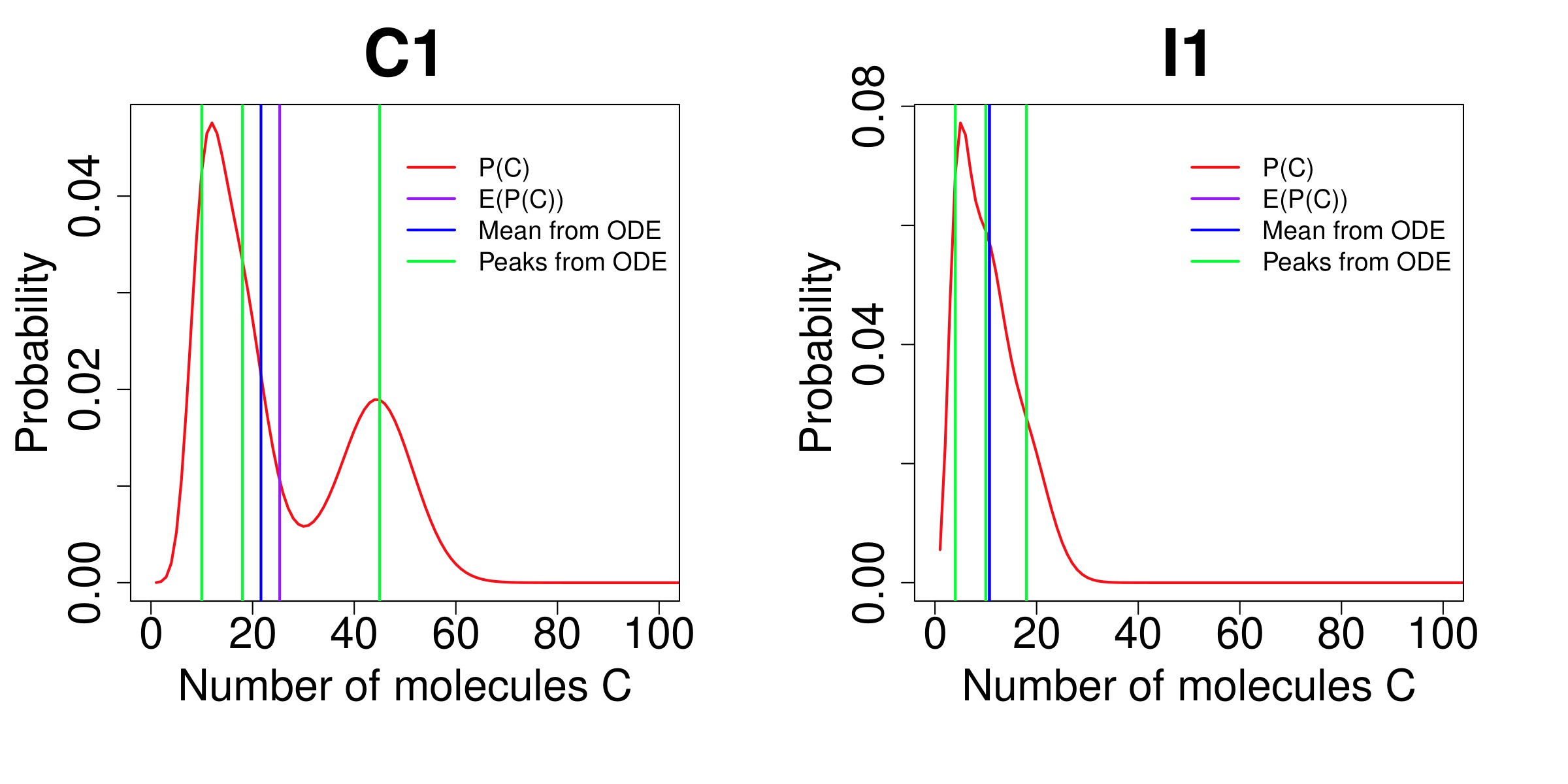}
\caption
{Comparing FFL behavior by ACME and by deterministic ODE models. 
(A) shows the results of FFL of C1 type for $(k_1, k_2, k_3)= (2.4, 4.5, 1.8)$. The exact results obtained using ACME exhibit bimodality in protein C (red curve), while trimodality is predicted by the deterministic ODE model (green vertical lines). The mean copy number from ACME (purple vertical line) is also different from the
that from ODE (blue vertical line). (B) shows the results of FFL of I1 type for $(k_1, k_2, k_3)= (2.4, 0.4, 1.8)$. 
The exact results obtained using ACME exhibit  monomodality in protein C (red curve), while deterministic ODE model predicts trimodality (green vertical lines), even though the mean copy number of protein C are the same between ACME and ODE models (purple and blue vertical lines, respectively). }
\label{fig:C}
\end{figure}

An example of the diverging results between ODE and stochastic models is shown in Fig.~\ref{fig:C}A
for an FFL of C1 type.
The mean values of $C$ obtained from the ODE model (vertical blue line) and the expectation computed from the probability landscape (vertical purple line) diverge from
each other~(Fig.~\ref{fig:C}A). There are three different phenotypic states by the ODE model (green lines, Fig.~\ref{fig:C}A), which are different from the bimodal probability 
distribution obtained from the SCK model~(Fig.~\ref{fig:C}A).

A further example is provided by the FFL of type I1.  
Here the ODE model predicts the existence of three phenotypes at $k_1=2.7, k_2=0.4$ and $k_3=1.8$~(Fig.~\ref{fig:C}B, green vertical lines). However, the stochastic model shows that there is only one stability peak. 
Although the mean value of $C$ obtained from the ODE model and the expected $C$ value computed from the probability landscape largely overlap, the ODE model provides no information on phenotypical variability.
Overall, stochastic models provide accurate and rich information that are not possible with ODE models.

\,

\subsubsection{Behavior of FFLs from exact solution to dCME by ACME can be differ from that by stochastic simulation algorithm}

\begin{figure}[h!]
\centering
\includegraphics[scale=0.35]{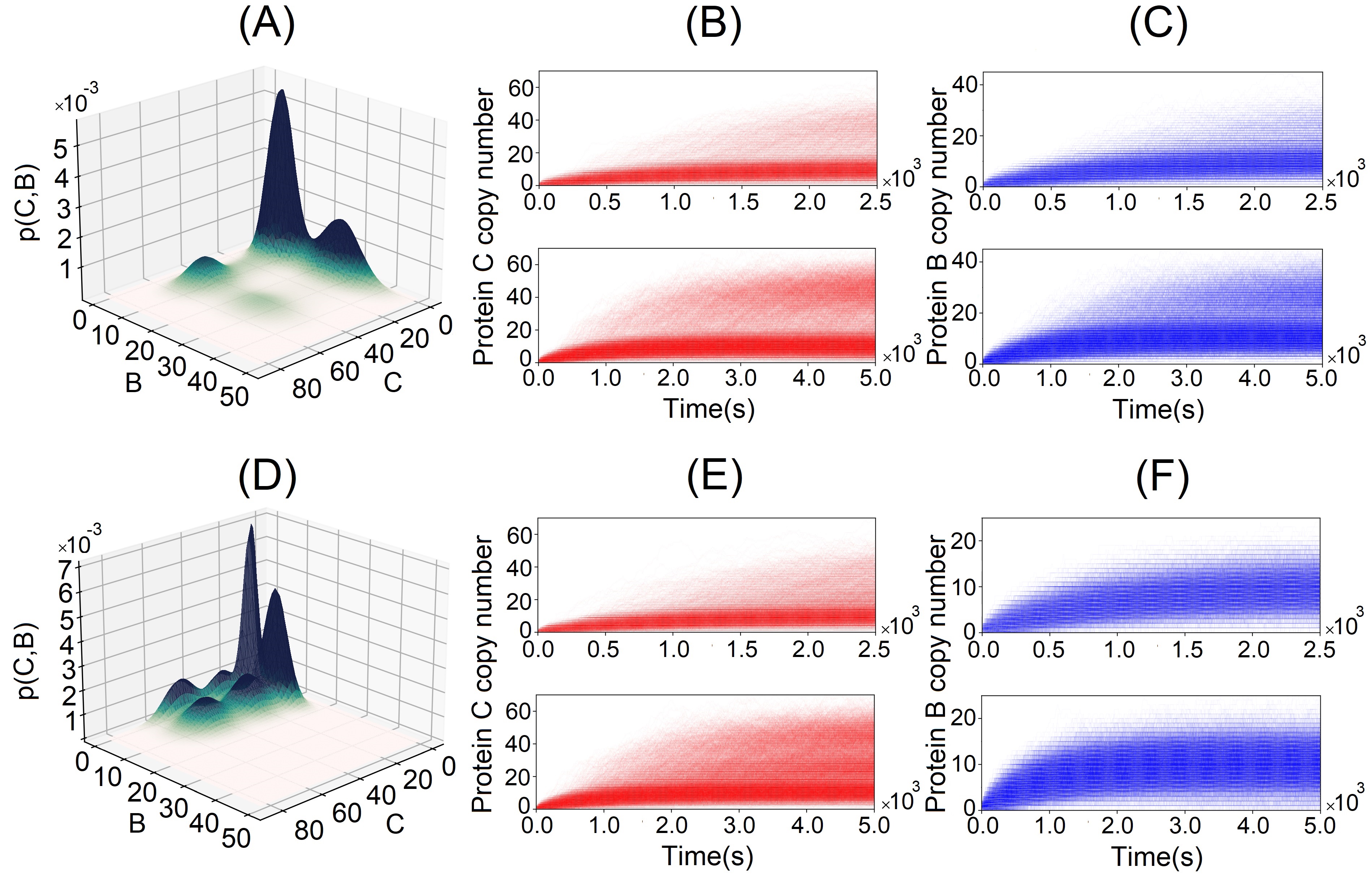}
\caption{Comparing landscapes from ACME and reaction trajectories from the stochastic simulation algorithm (SSA). (A) Probability surface projected onto the $(B,C)$-plane for the FFL with $(k_1,k_2,k_3)=(3.0, 0.5, 5.0)$. There is bimodality in both protein B and protein C. (B) and (C) The reaction trajectories computed from SSA corresponding to condition in (A) for protein C and protein B, respectively. The upper plots are for 2,500s and lower plots are for 5,000s. SSA does not capture the bimodality of protein B and C until 2,500s. (D) The probability surface projected onto $(B-C)$ plain for FFL with $(k_1,k_2,k_3)=(0.1, 2.75, 5.0)$. There is tri-modality in protein C and bimodality in protein B. (E) and (F) Corresponding reaction trajectories in protein C and protein B, respectively. Upper plots are for the results for 2,500s and lower plots are for 5,000s. SSA does not capture tri-modality of protein C until 2,500 seconds. In addition, SSA fails to capture bimodality in protein B.}
\label{fig:SSA}
\end{figure}

Results from 
simulations using SSA may differ from the exact solution to dCME obtained using ACME.
We illustrate this using two  incoherent FFLs, one at $(k_1,k_2,k_3)=(3.0, 0.5, 5.0)$ of I1-FFL
(Fig~\ref{fig:SSA}A-\ref{fig:SSA}C)
and another at $(k_1,k_2,k_3)=(0.1, 2.75, 5.0)$ (Fig~\ref{fig:SSA}D-\ref{fig:SSA}F) of the I4-type FFL. 
The exact steady-state probability landscape of the I1-FFL network computed  using ACME is
 multimodal,  exhibiting two peaks in protein B and two peaks in  protein C~(Fig.~\ref{fig:SSA}A).
However, these peaks are not definitive when 30,000 reaction trajectories up to 2,500 seconds  are simulated  using SSA (upper plots, Fig.~\ref{fig:SSA}B-\ref{fig:SSA}C).
Bimodality in protein B and protein C becomes only definitive when simulation time is extended to  5,000 seconds (lower plots, Fig.~\ref{fig:SSA}B-\ref{fig:SSA}C).

The exact steady-state probability landscape of the I4-FFL network
computed  using ACME 
exhibits tri-modality in protein C and bimodality in protein B~(Fig.~\ref{fig:SSA}D).
However, tri-modality
is not clearly captured when the reaction trajectories are $<2,500$ seconds  (upper plot, Fig.~\ref{fig:SSA}E), and becomes definitiveonly after 5,000 second (lower plot, Fig.~\ref{fig:SSA}E).
In addition, bimodality in protein B is not captured, even when the reaction trajectories are at $5,000$ seconds (upper and lower plot, Fig.~\ref{fig:SSA}F).

\,

\subsection{Phase diagrams of multimodality in FFLs.}
\begin{figure}[htb]
\centering
\includegraphics[width=1\linewidth]{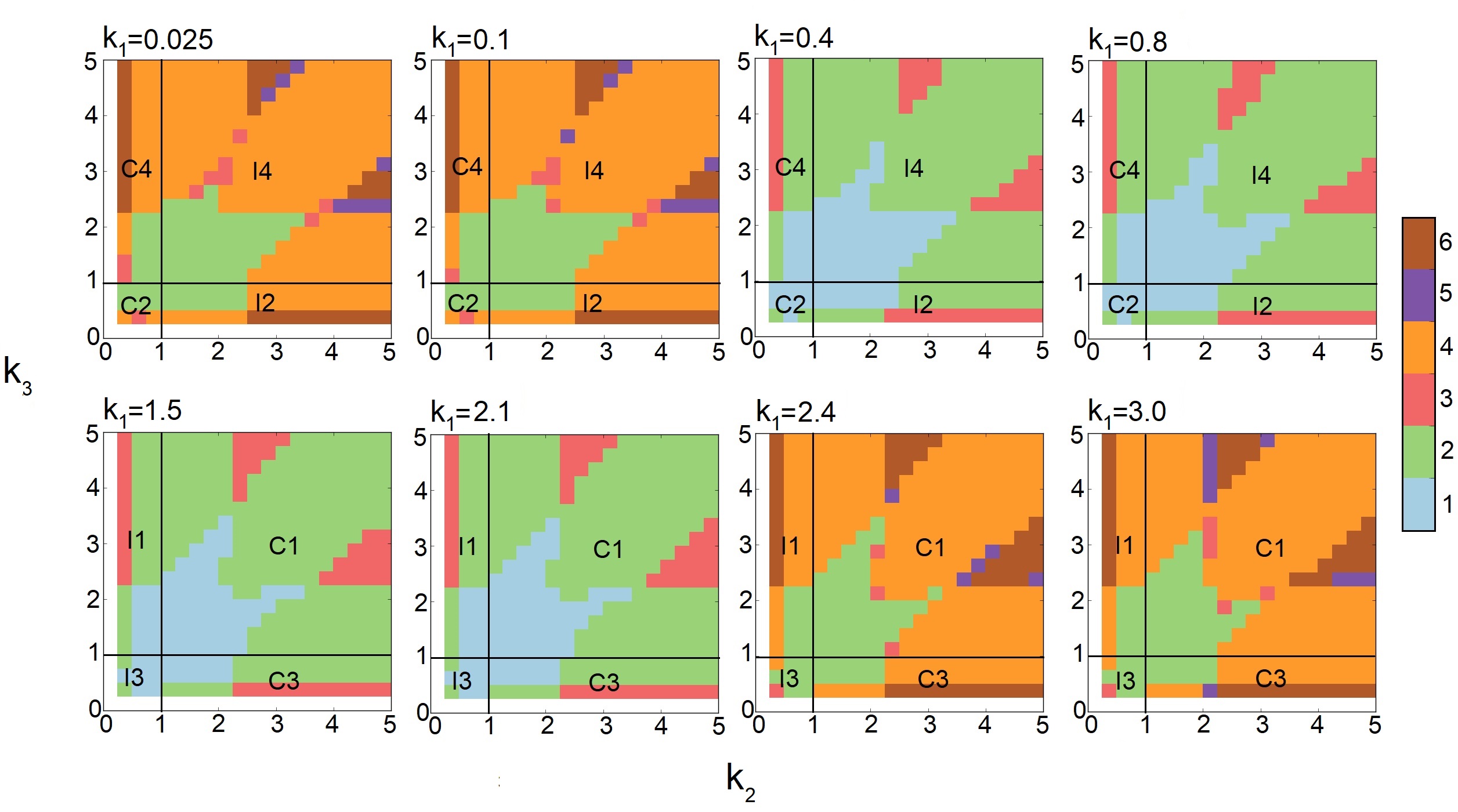}
\caption
{
Phase diagrams of multimodality of FFL network modules based on 10,812 steady state probability landscapes at different condition of regulation intensities  for all 8 types of FFL network modules. 
\textbf{Monomodality} occurs when $ 0.4 \le k_1\le 2.1$ and $k_2, k_3$ intensities are moderate, \textit{i.e.}, $ 0.4 \le k_1\le 3$ (blue region when $k_1=0.4, 0.8, 1.5,$ and $2.1$). \textbf{Bimodality} may occur for different combinations of regulation intensities. When $k_1$ intensity is either very high ($2.4 \leq k_1$) or very low ($k_1\leq 0.1$),
bimodality occurs when $k_2, k_3$ intensities are moderate, \textit{i.e.}, $ 0.4 \le k_1\le 3$. When $k_1$ intensity is moderate ($0.4 \le k_1 \le 2.1$), bimodality occurs when at least one of the other regulation
intensities $k_2$ or $k_3$ is high. \textbf{Tri-modality} occurs when $k_1$ is moderate ($0.4 \le k_1 \le 2.1$) and either  $k_2$ or  $k_3$ is moderate. \textbf{Multimodality} occurs when $k_1$ is low or high ($k_1 \le 0.4$ or $k_1 \ge 2.1$), and
at least
either $k_2$ or  $k_3$ 
is high.
Color scheme (vertical bar): Blue, green, red, orange, purple, and brow represent regions with one, two  three, four, five, and six peaks, respectively.
}
\label{fig:Fig_diagrams}
\end{figure}

Current studies of stochastic networks are limited to their behavior under a few selected conditions. Here we explore the multimodality  of all eight types of FFLs under broad conditions of synthesis, degradation, binding and unbinding as outlined in Table~\ref{tab:1}. This is made possible by the efficiency of the multi-finite buffer ACME algorithm. The analysis using persistent homology further allows us to quantitatively characterize the exact topology of the landscape. Together, we are able to obtain 
the full phase diagrams on the phenotype of multimodality of  FFLs at different combinations of
 parameter values~(Fig.~\ref{fig:Fig_diagrams}).

 Altogether, we compute  10,812  probability landscapes of the 8-types of FFL modules. 
Depending on the values of $k_1, k_2,$ and $k_3$, each phase diagram shown 
depicts the behavior of four types of FFLs, one for each of the four quadrants formed by 
the two straight lines of $k_2=1$ and $k_3=1$ (Fig.~\ref{fig:Fig_diagrams}), with the type of FFL labeled accordingly.  The specific types also depend on  $k_1$, which is listed at the top of each plot~(Fig.~\ref{fig:Fig_diagrams}).
As a result, we have gained comprehensive and accurate characterization of the multimodality phenotypes of this type of important network modules.

{\bf Monomodality.} As shown in Fig.~\ref{fig:Fig_examples}, the steady-state probability landscape of the FFL  at $k_1 = k_2 = k_3 = 1.2$ exhibits one probability peak. 
At this condition, it is a coherent FFL of type C1.
The projected distributions of $B$ and $C$ exhibit monomodality and has only one peak (Fig.~\ref{fig:Fig_examples},  red) when the values of intensities $k_1$, $k_2$, and $k_3$ are close to $1.0$ (Fig~\ref{fig:Fig_diagrams}).
Overall, there is only one phenotypic state when the regulations intensities in FFL are weak. 

{\bf Bimodality.} 
The steady-state probability landscape of FFLs can exhibit two types of bimodality
(colored yellow in Fig.~\ref{fig:Fig_loop}).
The first type occurs when $k_1 < 0.4$ or $k_1\geqslant 2.4$, with bimodality in protein $B$ while  monomodality in protein $C$. This is illustrated as green regions in Fig.~\ref{fig:Fig_diagrams} shown 
at the two top-left and the two bottom right phase diagrams where $k_1\in\{ 0.025,~ 0.1,~ 2.4,~ 3.0\}$.
That is, if the regulation intensities of $k_1$ and $k_2$ are about two fold different either way, bimodality in $B$ arises. 

The second type of bimodality occurs when  $0.4 \leq k_1 < 2.4$,
where protein $C$ exhibit bimodality while  monomodality is maintained in $B$.  This is illustrated as green regions in the remaining phase diagrams of Fig.~\ref{fig:Fig_diagrams}, where $k_1\in\{ 0.4,~ 0.8,~ 1.5,~ 2.1\}$.

{\bf Tri-modality.} 
The steady state probabilistic landscape of FFL can 
exhibit tri-modality (green, Fig.~\ref{fig:Fig_examples}). 
There are  three possible phenotypes in protein $C$ while monomodality in protein $B$ is maintained. 
Trimodal regions are colored red in the phase diagrams of Fig.~\ref{fig:Fig_diagrams}.
They arise when the difference in rates $k_2$ and $k_3$ is at least about two folds and $0.4 \leq k_1 \leq 2.1$.

{\bf Multimodality.} 
The steady state probability landscape of the FFL can exhibit 4 to 6 probability peaks (orange, purple, and green, respectively in Fig.~\ref{fig:Fig_examples}). 
Landscapes with 4 modes have bimodality in both protein $B$ and protein $C$. Those with 5 modes has bimodality in $B$ and tri-modality in $C$. Landscapes with 6 modes exhibit bimodality in $B$ and tri-modality in $C$.
Inspection on the conditions indicates that when the regulations are strong; i.e. when $k_1$, $k_2$, and $k_3$ $\ge 2.1$, FFLs exhibit very well defined multimodality peaks.
However, when the regulation intensity $k_1$ is weak, the steady state probability landscape exhibits multimodality only when the other  two  regulation intensities, namely, $k_2$ and $k_3$ are strong.
As shown in Fig.~\ref{fig:Fig_diagrams}, there are two groups of FFLs based on the characteristics of the multimodality they exhibit: 
One group consists of FFLs of $C_2$, $C_4$, $I_1$, and $I_3$ types, where tri-modality of output protein $C$ always exists,
as long as 
$k_2$ and $k_3$ are at least about two-fold different.
The other group consists of FFLs of $C_1$, $C_3$, $I_2$, and $I_4$ types where the signs of the regulations that the output node $C$ receives from $B$ and $A$ are the same (both activation or both inhibition).  
Tri-modality 
occurs when the regulations $k_2$ and $k_3$ have very distinct values.

Overall, protein $B$ can exhibit either mono- or bimodality, and protein $C$ can exhibit mono-, bi-, or tri-modality on the probability landscape.

\

\,

\subsection {Increasing input intensity amplifies multimodality in FFL}

To understand how input intensity affect the response  of FFL networks,  we examine their behavior under different
input conditions.  Specifically, we examine how different synthesis rate $s_A$ of protein A  affects the number of modes in proteins $B$ and $C$. 

\begin{figure}[h!]
\centering
\includegraphics[scale=0.60]{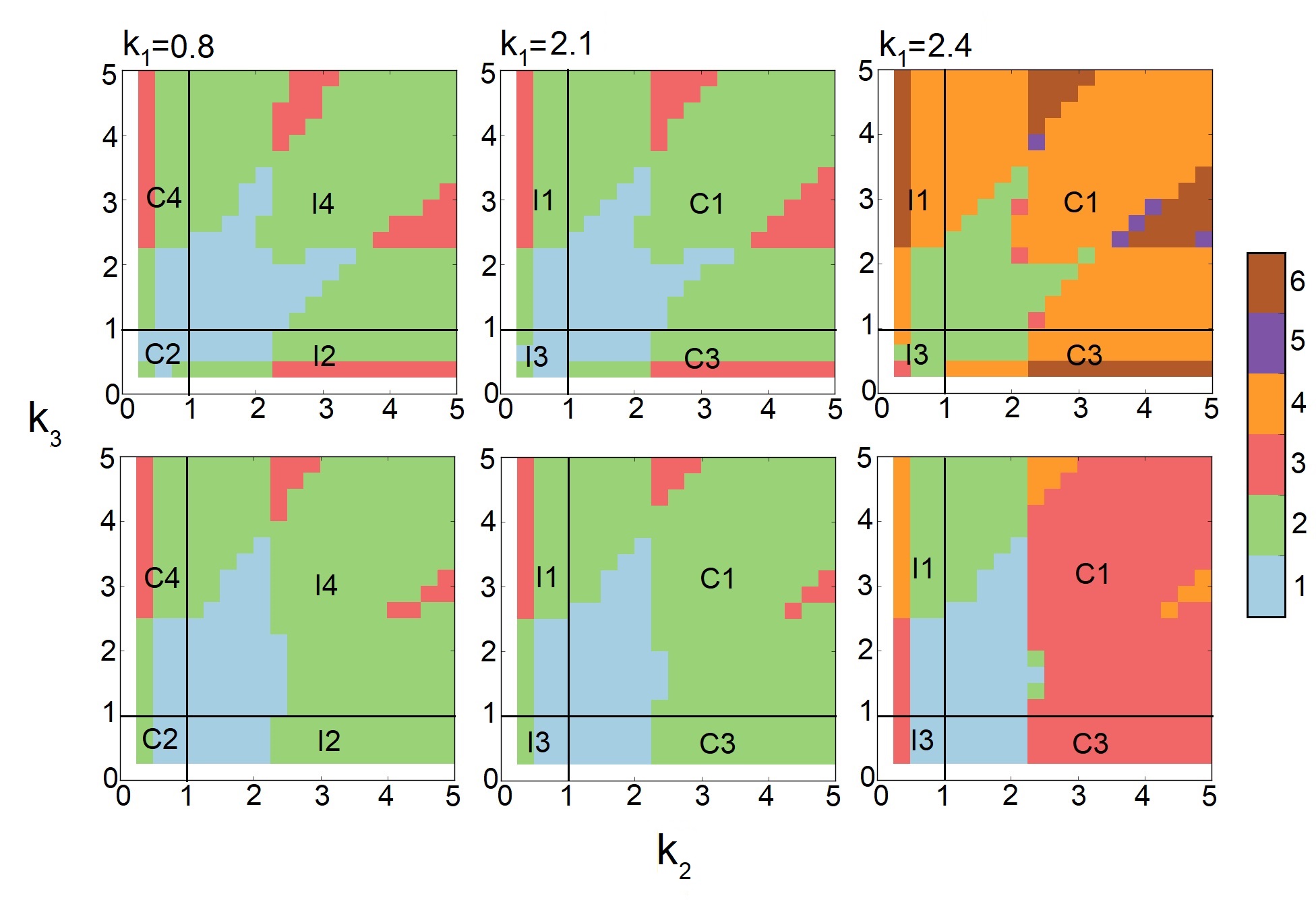}
\caption{Effects of input intensity on multimodality of FFLs. The phase diagrams of the number of stability peaks in the steady state probability landscapes at strong input intensity $s_A=10.0$ (\textit{top row}) and weak input intensity $s_A = 3.0$ (\textit{bottom row}) for different $k_2$ and $k_3$ at three different conditions of $k_1=0.8,\, 2.1,$ and $2.4$.
Color scheme (vertical bar): Blue, green, red, orange, purple, and brow represent regions with one, two,  three, four, five, and six peaks, respectively.
} 
\label{fig:Fig5}
\end{figure}

We first carry out computations and broadly survey the behavior of FFLs at strong input intensity, where $s_A$ is set to $s_A=10.0$.  The values of $k_2$ and $k_3$ are sampled broadly, and $k_1$ is tested for three different values of $k_1= 0.8, \, 2.1,$ and $2.4$.  The results are summarize in Fig.~\ref{fig:Fig5} (top row). We then similarly survey the behavior of FFLs at decreased synthesis intensity of protein A, with $s_A = 3.0$ (Fig.~\ref{fig:Fig5}, bottom row).

There are clear changes in the mode of multimodality of FFLs. At $k_1= 0.8$ and $k_1=2.1$ (Fig.~\ref{fig:Fig5}, left and center columns), when protein $A$ synthesis rate $s_A$ is reduced from $10.0$ (top) to $3.0$ (bottom), regions with one (blue) and three (red) peaks are reduced.
In addition, certain areas of the tri-stable (red) regions become bimodal (green).

At larger $k_1=2.4$ (Fig.~\ref{fig:Fig5}, right column), the FFLs exhibits dramatic changes in the modes of multimodality when synthesis rate $s_A$ of protein $A$ is reduced from 10.0 (top) to 3.0 (bottom). 
In many regions, one or more stability peaks disappear.
There are regions with two peaks at $s_A=10.0$ that become monomodal.  There are also regions of six peaks that become those of  four peaks. This is due to the loss of one stability peak from three in protein $C$ . In addition, large regions with four peaks (orange) disappear and become either regions with two peaks (green) or with three peaks (red).  
Overall, we can conclude that high input intensity represented by high $s_A$  rate for protein $A$ induces changed phenotypes of multimodality in FFLs.

\ 

\subsection {Binding and unbinding dynamics are critical for multiple phenotypic behavior}

Results  obtained so far are based on the assumption of slow  binding ($r_b^A = r_c^A = r_c^B = 0.005$) and unbinding ($f_b^A = f_c^A = f_c^B = 0.1$) reactions, which we call the \textit{generic case}. 
When the FFL network slowly switches between phenotypic states, the process of synthesis-degradation of protein $C$ has sufficient time to 
converge to equilibrium at each phenotypic state of gene $c$. An important questions is how slow the promoter dynamics need to be for FFLs to exhibit multiple phenotypes, without feed-back loops or cooperatively.

\begin{figure}[h!]
\centering
\includegraphics[scale=0.54]{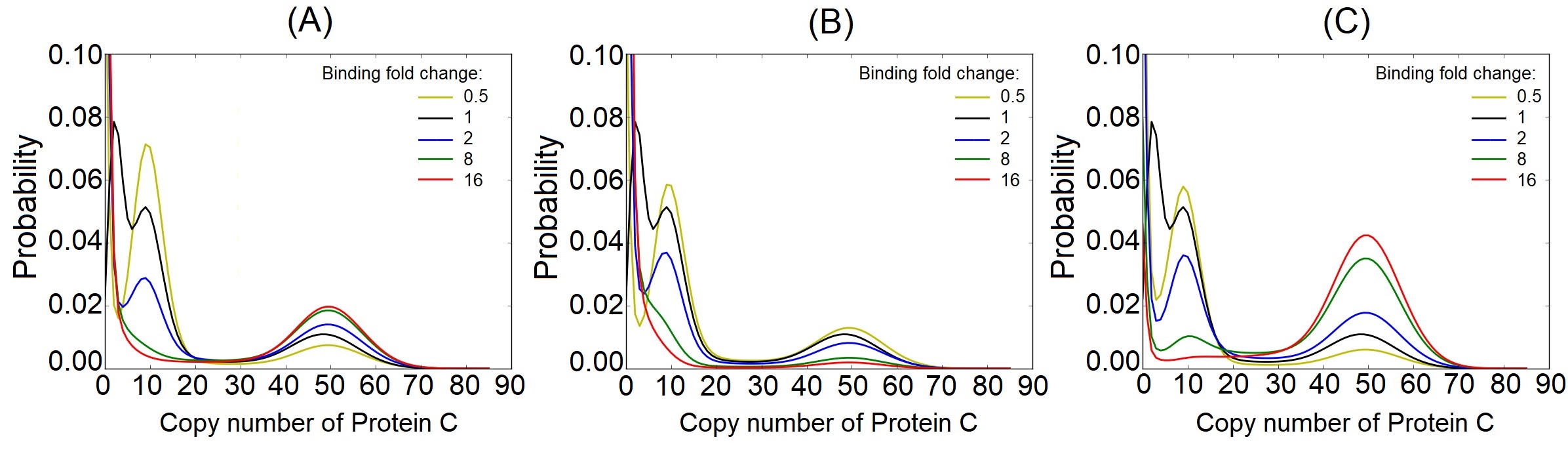}
\caption{Effect of binding dynamics on the modality of protein $C$ in the FFL network of type I1, with $(k_1,k_2,k_3)= (3.0, 0.025, 5.1)$. (A) Effects when binding affinity between  gene {\it c} and both protein $A$ and protein $B$ are altered by $n$-fold, where $n \in \{0.5,\, 2,\,  8,\, 16\}$. At slower binding (yellow line), the modes of distribution of protein C are well distinguished. However, when the binding and unbinding rates increased to 8 (green line), the peak at $C=9$ 
disappears. At $n=16$, 
bimodality is observed in protein $C$. (B) Effects when only the binding affinity of gene {\it c} and protein $B$ is 
altered by $n$-fold, where $n \in \{0.5,\, 2,\,  8,\, 16\}$. When the binding affinity of gen {\it c} and protein $B$ increases, 
the peak at $C=9$ disappears, while the peaks at $C=49$ robustly remains. However, the peak at $C=49$ becomes less significant. (C) Effects when only the binding affinity of gen {\it c} and protein $A$ is altered by $n$-fold, where $n \in \{0.5,\, 2,\,  8,\, 16\}$. At high binding affinity, the peak at $C=9$ disappears while the peak at $C=49$ becomes more prominent. }
\label{fig:Fig6}
\end{figure}

To answer this question, we explore the behavior of FFLs under different binding and unbinding dynamics of gene $c$ for a FFL of type I1.
In this case, protein $A$ activates protein $B$ and protein $C$, while protein $B$ inhibits protein $C$ (see Fig.~\ref{fig:Fig_loop}B). 
With slow binding kinetics as described above, the output $C$ of this FFL  exhibits three stability peaks. These are at the expression level of protein $C$ of
1) $C=0$, corresponding to the condition  when gene $c$ is inhibited by $B$; 2) $C=9$, corresponding to the basal level of $C$ expression,
and 3) $C=49$, when $C$ expression is activated by $A$.
We then fix the regulation
intensities at $k_1=3.0$, $k_2=0.025$,  and $k_3=5.1$, and examine how the number of phenotypic states is affected by  gene $c$ binding dynamics (Fig.~\ref{fig:Fig6}).

We first set the binding affinities between gene $c$ and protein $A$ and between gene $c$ and protein $B$ 
to the same values, and change them together to $n$-fold of the \textit{generic case}, where $n \in \{0.5,\, 2,\,  8,\, 16\}$.
For slower binding and unbinding dynamics (yellow line for $n=0.5$, Fig.~\ref{fig:Fig6}A), the modes of the distribution of the output of protein $C$ are even better distinguished.
However, when both binding and unbinding rates are increased to $n=8$ fold (green line), the probability peak at $C=9$, which corresponds to basal level of C expression, merges with the probability peak at $C=0$.
At $n=16$,  the distribution of $C$ is bimodal.

We then keep the biding affinity between gene $c$ and protein $A$ unchanged and alter only the binding affinity between gene $c$ and protein $B$ by $n$-fold, where $n \in \{0.5,\, 2,\, 8,\, 16\}$.
 When the binding affinity increases ({\it e.g.}, $n=8$), the probability peak at $C=9$ disappears, while the probability peak at high copy number of $C=49$ robustly remains, although with less magnitude~(Fig.~\ref{fig:Fig6}B).

When only the biding affinity between gene $c$ and protein $A$ is altered while that between gene $c$ and protein $B$ is held constant (Fig.~\ref{fig:Fig6}C), the probability peak at the basal level of C expression ($C=9$) diminishes when the binding affinity increases ({\it e.g.}, $n=8$).  However, the probability peak at $C=49$ becomes more prominent.
At $n=8$, the distribution of C is tri-modal. At  $n=16$, it becomes bimodal.  This suggests that multiple phenotypes arise in FFLs when the unbinding rate is about an order of magnitude smaller than the expression rate of the protein.
 
\

\subsection {Gene duplication can enrich phenotypic diversity and enlarge stable regions of specific multimodality of FFLs}

Gene duplication provides a basic route of evolution~\citep{lynch2000evolutionary} and is an important driver of phenotypical diversity in organisms~\citep{conrad2007gene}. 
Here we study how gene duplication affects the phenotypes of FFLs.

We examine how duplication of gene $c$, and separately  duplication of  gene $b$, affect the behavior of the FFL network modules. 
With two copies of gene $c$, there can be six possible states of gene $c$ activation. Depending on whether the promoter sites of both copies of gene $c$ are free or occupied by either protein $A$ or protein $B$, we have for both  $c$ genes to have unoccupied, protein $A$ bound, or protein $B$ bound promoter site.  This can be denoted as a triplet $(c,  cA, cB)$, which can take any of the
possible values of $(2, 0, 0)$, $(0, 2, 0)$, $(0, 0, 2)$,  $(1, 1, 0)$, $(1, 0, 1)$, and $(0, 1, 1)$. 
For the case when there are two copy number of gene $B$, there are three
possible states of gene $b$ activation,
depending on whether the promoter site of both copies of gene $b$ are free or occupied by protein $A$. This can be denoted as a duplicate $(b, bA)$, which can take any of the possible values of $(2, 0)$, $(1, 1)$, or $(0, 2)$.

\begin{figure}[h!]
\centering
\includegraphics[scale=0.49]{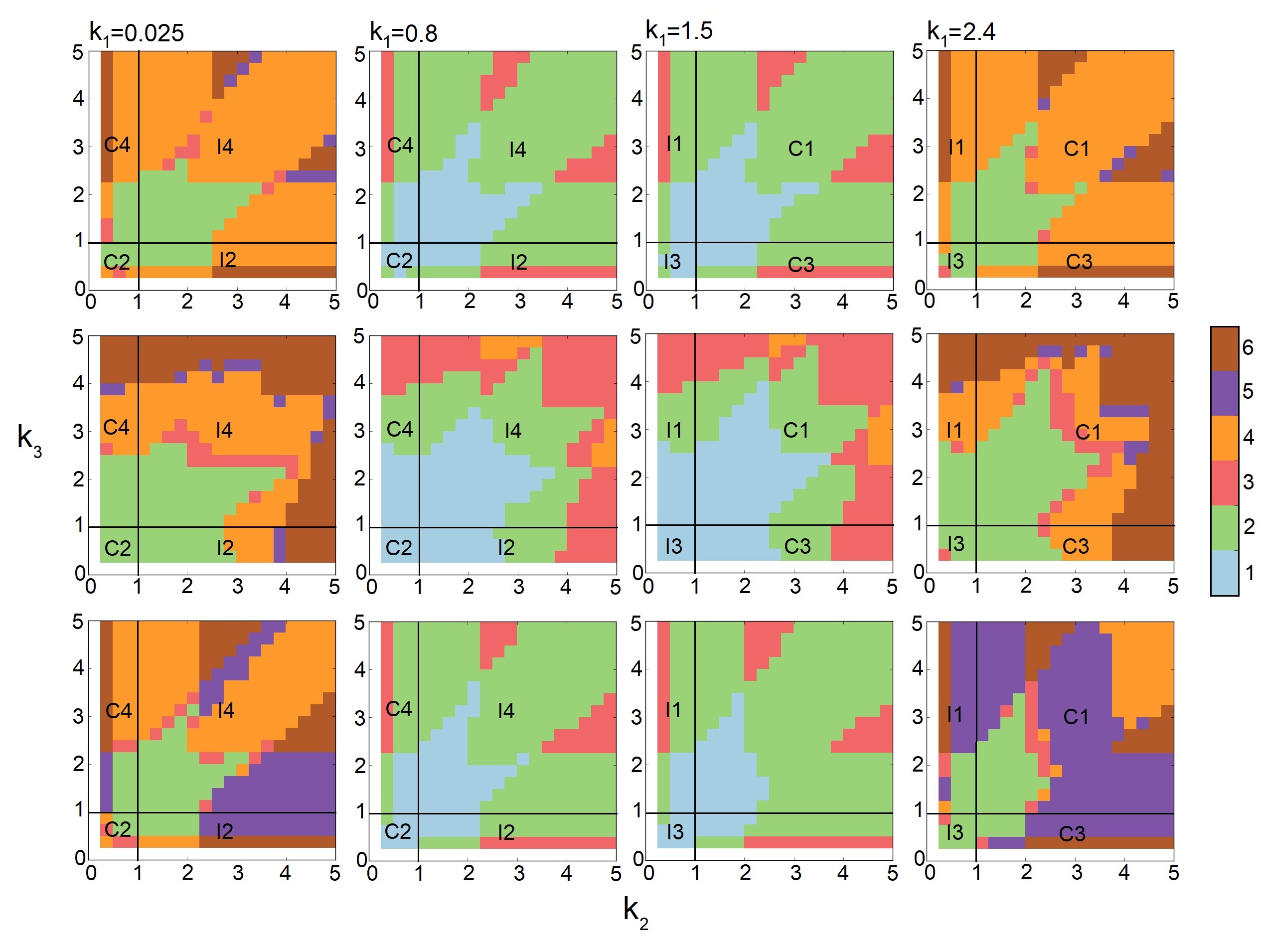}
\caption{ Phase diagram of the effects of gene duplication on multimodality of FFLs. 
(\textit{First row})  
Phase diagrams of 
the modality of stability peaks when there are one copy of gene $c$ and one copy of gene $b$. 
(\textit{Second row})  Phase diagrams when there are one copy of gene $b$ and two copies of gene $c$. 
(\textit{Third row}) Phase diagrams 
when there are two copy  of gene $b$ and one copy of gene $c$. The first, second, and third columns are for $k_1= 0.025,\, 1.5$ and $2.4$, respectively.
Color scheme (vertical bar): Blue, green, red, orange, purple, and brown represent regions with one, two  three, four, five, and six peaks, respectively.}  
\label{fig:Fig4}
\end{figure}

The phase diagrams of the number of modes of stability peaks are shown in Fig.~\ref{fig:Fig4},   
when there is only one copy  of both gene $b$ and gene $c$ (first row), when there are two copies of gene $c$ but one copy of gene $b$ (second row), and when there are two copy number of gene $b$ but one copy of gene $c$ (third row).  The conditions are $k_1=0.025,\, 0.8,\, 1.5,$ and $2.4$, for different values of $k_2 \in [0.1,\, 5]$ and $k_3 \in [0.1,\,5]$, where there are slow binding and unbinding ($r_b^A = r_c^A = r_c^B = 0.005$, $f_b^A = f_c^A = f_c^B = 0.1$). 
Each phase diagram in Fig.~\ref{fig:Fig4} consists of $400$ steady-state probability landscapes with a total $12 \times 400 = 4,800$ landscapes. This broad range of parameters allow us to study all 8 different modules of FFL network and the effects of gene $c$ and gene $b$  duplications.

We examine the behavior of FFL in three different regimes of $k_1$: 1) When $k_1 \ll 1.0$ (Fig.~\ref{fig:Fig4}, first column), the  bimodal regions (green) expands  when there are two copies of gene $c$ (second row), but there are no significant changes when there are two copies of gene $b$ (third row). 
In addition, 
the overall size of multimodal regions increases in both cases;
2) When $k_1 \approx 1.0$  (Fig.~\ref{fig:Fig4}, second and third columns), 
the duplication of gene $c$ (second row)
expands the regions with three stability peaks and reduces regions with two peaks. In contrast,
the duplication of gene $b$ (third row) has no significant effects on  multimodality; 
3) When $k_1=2.4$ (fourth column), duplication of gene $c$ (second row) expands regions
with two and six stability peaks.
Duplication of gene $b$  (third row) reduces the region with four peaks and expands the region with five peaks.

These results show that introducing additional copy of gene $b$ or gene $c$   can not only enrich  different phenotypic behavior, but also increase the stability of specific phenotypic states, namely, enlarge regions of particular phenotypes by uniting previously different phenotypic regions together.  Overall, gene duplication can increase phenotypic diversity, and enlarge stability regions of specific multimodal states.

Bacterial cells have fast binding and unbinding dynamics~\citep{bacteriaFastBindingUnbinding}, it is unlikely that the occurrence of multiple copies of the same gene in FFLs play significant roles in stochastic multimodality. 
In contrast, mammalian cells have slower promoter 
dynamics~\citep{slowerBindingUnbindingDynamics}. Gene duplication in FFLs may provide a natural mechanism for enriched multimodality with enhanced stochastic phenotypic switching.
This is reflected in reduced monomodal regions, and enlarged multimodal regions where there are 4 (orange), 5 (purple), and 6 (brown) phenotypic states  of the output $C$ (second and third row on Fig.~\ref{fig:Fig4})

Assuming that initially both copies of the gene were functioning, but subsequently one gene copy lost its biochemical function due to 
mutations, we can expect two opposite types of scenarios to occur: 
If regulation intensities are strong ($k_2$ and $k_3$ are large), 
one of the phenotypic states becomes lost (\textit{e.g.}, green region becomes light blue, and orange region becomes red, Fig.~\ref{fig:Fig4}).
If regulation intensities are weak,
the duplication of gene $c$ or gene $b$ can lead to enlargement of the region of monomodality. It can also lead to the appearance of new regimes where there are a larger number of multimodality modes (orange, purple, and green regions in Fig.~\ref{fig:Fig4}). 
That is, gene duplication can create new stable states, leading to an enlarged number of high probability states. This, however, occurs only in FFL modules with 
strong regulations intensities.  FFL modules
with low regulation intensities instead lose phenotypical diversity and become more robust in monomodality with enlarged region in the parameter space.

\section {Discussion}

Gene regulatory networks (GRNs) play critical roles in defining cellular phenotypes but it is challenging to
characterize the  behavior of GRNs. 
Although GRNs may consist of dozens or more of genes and proteins, their functions often can be defined by smaller sub-networks called network motifs. 
How small network motifs are responsible for complex
properties such as the maintenance of multi-phenotypic behavior or modules is poorly understood.
Current widely practiced approach is studying network motifs in that of deterministic models. However, this approach imposes restrictions on the types of network motifs capable of exhibiting multimodal phenotype  to mostly feed-back networks.

In this study, we examined the feed-forward loop network motifs, one of the most ubiquitous three-node network motifs.
Although their deterministic  behavior is well studied, with great understanding of their functions such as signal processing and adaptations, their stochastic behavior remains poorly characterized.

Here, we showed the direct regulation path from the input node to the output node, and the indirect path through the intermediate buffer node provide the necessary architecture for  distinct multiple modalities. Phase diagrams of FFL in Fig.~\ref{fig:Fig_diagrams} show that FFLs of various types can exhibit different multimodality. At large copy numbers and large volume, our model of stochastic reaction kinetics are the same as those based on mass action kinetics~\citep{kurtz_1971,kurtz_1972,QianHong2007}, where ordinary differential equation (ODE) models are appropriate.  When ODE models are applied to enzyme-substrate reactions, they can be approximated by Michaelis-Menten kinetics,  with the additional assumption that
 the substrate is in instantaneous chemical equilibrium with the enzyme-substrate complex. 
 When ODE models are applied to
 the reaction of one receptor and $n$ identical
 simultaneously binding ligands, we arrive at the Hill equation, with the coefficient $n$  phenomenologically characterizing cooperativity.
These kinetic models based on ODE approximations, however, are not applicable to the current study, as we are examining strong stochasticity arising at low copy number of molecules, where ODE models are not valid.

FFLs play important roles in gene regulatory networks. 
For example, it is shown that  several I1-FFL sub-networks control the process of  \textit{Bacillus subtilis sporulation}~\citep{Eichenberger2004,mangan2006incoherent}. 
In addition, C1-FFL  network is found to be present in the \textit{L-arabinose} (\textit{ara}) utilization system of \textit{E. coli}, where \textit{araBAD} is the
target (gene {\it c}) activated by the intermediate gene \textit{araC} and the input gene \textit{CRP}. 
Gene \textit{araC} is also activated by \textit{CRP}. Therefore, they form  a 3-node C1 type FFL~\citep{MANGAN2003197}. 
Results in this work can help to gain understanding of the behavior of these different types of FFLs found in gene regulatory networks.

In addition, we have shown that input intensity affects the multimodal behavior of various types of FFLs. Examples shown in Fig.~\ref{fig:Fig5} demonstrate
that at high $k_1$ values, input intensity dramatically changes the phase diagram. 
Our results are consistent with previous findings that
input intensity is an
important factor in determining output intensity of FFLs~\citep{MANGAN2003197, GOENTORO2009894, LinYen-Potoyan2018}. Here we further demonstrated that
input intensity is also
important in determining the modality of the steady state behavior of FFLs.

In mammalian cells,  slow dynamics of transcription factor binding to promoter is
often observed~\citep{Hasegawa2019,lickwar2012genome,HAGER2009741,Dermitzakis2002,tugrul2015}.  This is likely due to the complex process of chromatin regions opening up so they become accessible  and the slow nature of events such as promoter, enhancer and mediator binding. These physical processes  result in   
highly stochastic behavior of networks.  
Stochastic models have demonstrated that complex multimodality phenotypes can naturally arise from stochastic fluctuations when genes have distinct expression levels,  
a phenomenon widely observed in mammalian cells~\citep{cao2018probabilistic}. 
We showed that binding and unbinding dynamics are critical for multi-phenotypic behavior. For an I1-FFL with $(k_1, k_2, k_3)=
(3.0,\, 0.025,\, 5.1)$, 
Fig.~\ref{fig:Fig6} highlighted that binding and unbinding rates affect multiple peaks in protein C.

Results of this study indeed showed that once stochastic fluctuations between distinct expression levels due to slow promoter dynamics are considered, FFLs can  exhibit complex multimodal phenotypes.
When the 
  expression levels of the output gene (gene $c$) at the inhibited, basal, and activated states are well separated, three distinct phenotypes arise.  Combined with two additional possible
phenotypes of different levels of gene $b$ expression, we can have up to six modalities for FFLs. 
Furthermore,  high intensity of input amplifies multimodality in FFLs, 
suggesting that the FFL architecture are favored for maintaining multiple phenotypic states.
In addition, we find that regulation intensities are key determinants of specific stochastic behavior of FFLs, which 
could be tuned in order to obtain any desired phenotypic behavior between 1 to 6 stability modes. 

Our study also revealed the roles of gene duplication.
 When there are two copies of gene $c$, while one in principal could expect $2\times 6=12$ different phenotypes  for the output protein $C$, this is, however, not observed, as the regulation intensities or reaction rates are not so well separated.
 In contrast, instead of further increase in in multimodality beyond six, 
we observe the expansion of the area of monomodality,  resulting from the connectedness of regions of expression with different rates that become  merged together.
  Our results showed that duplication of gene $b$ and gene $c$ not only can enrich different phenotypic behavior, but can also increase the stability of certain phenotypic states, while decreasing others (Fig.~\ref{fig:Fig4}). We showed that in general, gene duplication can enrich phenotypic diversity. 
The presence and functional roles of gene duplication are well-known~\citep{Hurles2004}. 
For example, in human induced pluripotent stem cells (HiPSCs), chromosome 12 duplication lead to significant enrichment of cell cycle related genes~\citep{MAYSHAR2010521}, in which FFL sub-networks play important roles. This abnormality results in 
increase in 
the tumorigenicity of HiPSCs. Our findings may also shed light on how gene duplication affects cellular adaptation to changing environment~\citep{kondrashov2012gene}: As the support regions of monomodality are enlarged, smaller fluctuations in regulation intensities  will not switch cells with duplicated genes to a different phenotypic state. Thus, gene duplication may help to stabilize the behavior of the network, so cells  are better adapted to a changing environment.

Analysis of stochastic behavior of FFLs reported here
have implications in a
variety of biological problems.
For example,
the stem cell regulation network
consisting of pluripotency transcription 
factors \textit{Oct4} and \textit{Nanog}  maintain pluripotency
against
differentiation~\citep{papatsenko2015single, BOYER2005947,chickarmane2006,LinYen-Potoyan2018}. A component of this network can be abstracted as an FFL: \textit{Nanog} participates as the intermediate node (gene \textit{b}, which is activated by \textit{Oct4} (gene \textit{a}), and both regulate the expression of genes associated with the onset of differentiation or pluripotency (gene \textit{c}s).
In addition, regulation networks 
in hematopoietic stem cells are formed by
 two FFL networks involving
  \textit{$\beta$~globin}, \textit{GATA-$1$}, \textit{EKLF}, and \textit{FOG-$1$}.
  In each network, \textit{FOG-$1$} and \textit{EKLF} function as the intermediate genes (gene \textit{b}), and are activated by \textit{GATA-$1$} (gene \textit{a}),  while all of them activate \textit{$\beta$~globin} (gene \textit{c})~\citep{SWIERS2006525}.
    Moreover, in other stem cell differentiation networks, there are 
    several sub-networks that exhibit behaviors of different types of FFLs. For example,  \textit{Klf4} (gene \textit{a}) activates \textit{Pou5f1} (gene \textit{b}) and inhibits \textit{Sox2} (gene \textit{c}), while \textit{Pou5f1} activated \textit{ Sox2}~\citep{OKAWA2016307, Onochtchouk2010}, as in the C3-type FFL (Fig.~\ref{fig:Fig_loop}).

In summary, we have constructed and analyzed  the exact high-dimensional steady state probability landscapes of FFLs under broad conditions and have constructed their phase diagrams in multimodality.
These results are based on 10,812 exactly computed probability landscapes and their topological features as measured by persistent homology.
With slow binding and unbinding dynamics of transcription factor binding to promoter,  FFLs exhibit strong stochastic behavior that is very different from deterministic models, and can exhibit from 1 up to 6 stability peaks.  
In addition, input intensity play major roles in the phenotypes of FFLs: At weak input intensity, FFL exhibit monomodality, but strong input intensity  may result in up to 6 stable phenotypes. 
Furthermore, we found that gene duplication can enrich the diversity of FFL network phenotypes and enlarge stable regions of specific multimodalities. 

Results reported here 
can be useful for constructing synthetic networks, 
and for selecting model parameters so a particular desirable phenotypic behavior can materialize~\citep{jones2020}.
Our results can also be used for analysis of behavior of 
feed-forward loops in biological processes such as stem cell differentiation and for design of synthetic networks with desired phenotype behavior.
We hope results reported here for different types of FFL can be tested experimentally.

\section*{Conflict of Interest Statement} 
There are no conflict of interests.

\section*{Author Contributions}
A.T. and J.L. conceived and designed the study.
A.T. designed and carried out analysis of the ODE model, multimodality, phase diagrams, slow dynamics, and gene duplication. F.M. designed and carried out analysis of persistent homology, and assisted in multimodality and phase diagram computation.
Y.C. participates in design and data analysis. A.T. and J.L. wrote the manuscript with significant input from F.M.
All authors read and approved the final manuscript.

\section*{Funding}
This work is supported by NIH grant R35 GM127084.

\section*{Acknowledgments}
We thank Wei Tian for assistance in computing persistent homology.

\bibliographystyle{frontiersinSCNS_ENG_HUMS} 

\bibliography{references}

\,

{\bf TABLES}

\begin{table}[h!]
    \centering
    \captionsetup{justification=centering}
    \caption{Parameter ranges for eight types of FFL model}
  \begin{tabular}{ | c | c | c | c |}
    \hline
    FFL Type & $k_1$ range & $k_2$ range & $k_3$ range \\ [0.5ex]
\hline \hline
$C_1$ & (1.0 ~ 3.0] & (1.0 ~ 5.0] & (1.0 ~ 5.0] \\
\hline
$C_2$ & [0.025 ~ 1.0) & (1.0 ~ 5.0] & (0.025 ~ 1.0] \\
\hline
$C_3$ & (1.0 ~ 3.0]  & [0.025 ~ 1.0) & [0.025 ~ 1.0)\\
\hline
$C_4$ & [0.025 ~ 1.0) & [0.025 ~ 1.0) & (1.0 ~ 5.0] \\
\hline
$I_1$ & (1.0 ~ 3.0]  & [0.025 ~ 1.0) & (1.0 ~ 5.0] \\
\hline
$I_2$ & [0.025 ~ 1.0) & [0.025 ~ 1.0) & [0.025 ~ 1.0)\\
\hline
$I_3$ & (1.0 ~ 3.0] & (1.0 ~ 5.0] & [0.025 ~ 1.0) \\ 
\hline
$I_4$ & [0.025 ~ 1.0) & (1.0 ~ 5.0] & (1.0 ~ 5.0] \\[1ex] 
\hline 
\end{tabular}
\label{tab:1}
\end{table}

\newpage

\section*{SUPPORTING INFORMATION}

\,

\subsection*{Integer equations approach to modeling of Feed-forward loop}

The equations governing the kinetics in the constructed feed-forward loop were developed as follows:
$$
\frac{{d[b]}}{{dt}} = u_b^{A} [bA] - b_b^{A} [A][b];
$$
$$
\frac{{d[bA]}}{{dt}} =b_b^{A} [A][b] -  u_b^{A} [bA];
$$
$$
\frac{{d[c]}}{{dt}} = u_c^{A} [cA] + b_c^{B} [cB] - b_c^{A} [A][c]-  b_c^{B} [B][c];
$$
$$
\frac{{d[cA]}}{{dt}} = b_c^{A} [A][c]-u_c^{A} [cA];
$$
$$
\frac{{d[cA]}}{{dt}} = u_c^{A} [B][c]-b_c^{A} [cB];
$$
$$
\frac{{d[A]}}{{dt}} = s_{A} -d_{A} [A];
$$
$$
\frac{{d[B]}}{{dt}} = s_{B} [b] + k_1*s_{B} [bA]-d_{B} [B];
$$
$$
\frac{{d[C]}}{{dt}} = s_{C} [c] + k_2 * s_{C} [cA]+k_3*s_{C} [cB]- d_{C}[C].
$$
Here $[A]$, $[B]$, $[C]$ are the concentrations of proteins $A$, $B$, and $C$ respectfully.
$[b]$, $[c]$ are the concentrations of genes $b$, and $c$,
and $[bA]$, $[cA]$, $[cB]$ are the concentrations of genes $b$, and $c$ in corresponding bound states $bA$, $cA$, $cB$.

Let the total amount of molecules of gene $b$ in the system be $n_b$, 
and total amount of molecules of gene $c$ in the system be correspondingly $n_c$. 
Assuming that the copy numbers of genes $b$ and $a$ are large, 
then the unique deterministic solution of the system of ordinary differential equations above are:

$$
[b]=\frac{{u_b^A n_b }}{{u_b^A  + b_b^A [A]}};
$$
$$
[bA]=\frac{{n_b b_b^A [A]}}{{u_b^A  + b_b^A [A]}};
$$
$$
[c]=\frac{{u_c^A u_c^B n_C }}{{u_c^A u_c^B  + u_c^B b_c^A [A] + u_c^A b_c^B [B]}};
$$
$$
[cB]=\frac{{u_c^A b_c^B n_C [B]}}{{u_c^A u_c^B  + u_c^B b_c^A [A] + u_c^A b_c^B [B]}};
$$
$$
[cA]=\frac{{u_c^B b_c^A n_C [A]}}{{u_c^A u_c^B  + u_c^B b_c^A [A] + u_c^A b_c^B [B]}}
$$
$$
[A]=s_A/d_A;
$$
$$
[B]=\frac{{s_b n_b }}{{d_b }}  \cdot  \frac{{u_b^A  + k_1 b_b^A [A]}}{{u_b^A  + b_b^A [A]}}
$$
$$
[C]=\frac{{s_C n_C }}{{d_C }} \cdot \frac{{u_c^A u_c^B  + k_2 u_c^B b_c^A [A] + k_3 u_c^A b_c^B [B]}}{{u_c^A u_c^B  + u_c^B b_c^A [A] + u_c^A b_c^B [B]}}
$$

However the copy numbers of genes in the real systems is small, and usually does not exceed two copies.
Thus the formulation of mass-action kinetics equation for $[b]$ and $[c]$ in a canonical way is not justified.
Indeed in the case of $n_b=n_c=1$, quantities $[b]$, $[bA]$, $[c]$, $[cA]$, and $[cB]$ are fractions in the interval $[0,\,1]$ for ODEs, whereas
they can only be equal to $0$ and $1$.
In this case, one can compute six integer steady state solutions with respect to  $[b]$, $[bA]$, $[c]$, $[cA]$, and $[cB]$ concentrations.
Every each of these solutions corresponds  
to particular combination of gene copy numbers. 
They are 8-tuples $([A], [B], [C], [b], [bA], [c], [cA])$, such as
$$(s_A/d_A, s_B/d_B, s_C/d_C, 1, 0, 1, 0, 0),$$
$$(s_A/d_A, k_1 s_B/d_B, s_C/d_C, 0, 1, 1, 0, 0),$$
$$(s_A/d_A, s_B/d_B, k_2 s_C/d_C, 1, 0, 0, 1, 0),$$
$$(s_A/d_A, k_1 s_B/d_B, k_2 s_C/d_C, 0, 1, 0, 1, 0),$$
$$(s_A/d_A, s_B/d_B, k_3 s_C/d_C, 1, 0, 0, 0, 1),$$
$$(s_A/d_A, k_1 s_B/d_B, k_3 s_C/d_C, 0, 1, 0, 0, 1).$$
From this deterministic assumption of a
discrete set of gene occupancies, we obtain six stable peaks for the system. However, as stated in the main text, when the rates of regulation are not  well separated,  these peaks are merged together.
Note that this approach is different from boolean modeling, where the analysis is conducted 
in a qualitative way (as ON/OFF output)~\citep{wang2012boolean}.

\end{document}